\documentclass[twocolumn,showpacs,pra]{revtex4}
\usepackage{graphicx}
\usepackage{dcolumn}
\usepackage{bm}

\begin{document}

\title{Quantum Jumps between Macroscopic Quantum States of a
Superconducting Qubit Coupled to a Microscopic Two-Level System}
\title{Quantum Jumps between Macroscopic Quantum States of a Superconducting
Qubit Coupled to a Microscopic Two-Level System}
\author{Yang Yu$^{1,2}$}
\email{ yuyang@nju.edu.cn}
\author{ Shi-Liang Zhu$^{3}$}
\email{slzhu@scnu.edu.cn}
\author{ Guozhu Sun$^{2}$}
\email{gzsun@nju.edu.cn}
\author{Xueda Wen$^{1}, $ Ning Dong$^{1}$, Jian Chen$^{2}$%
, Peiheng Wu$^{2}$}
\author{Siyuan Han$^{2,4}$}
\affiliation{$^{1}$National Laboratory of Solid State
Microstructures and Department of
Physics, Nanjing University, Nanjing, China\\
$^{2}$Department of Electronic Science and Engineering and RISE,
Nanjing
University, Nanjing, China\\
$^{3}$Institute for Condensed Matter Physics and SPTE, South China
Normal
University,Guangzhou,China \\
$^{4}$Department of Physics and Astronomy, University of
Kansas,Lawrence, KS
66045, USA}
\begin{abstract}
We report the observation of quantum jumps between macroscopic quantum
states in a superconducting phase qubit coupled to the two-level systems in
the Josephson tunnel junction, and all key features of quantum jumps are
confirmed in the experiments. Moreover, quantum jumps can be used to
calibrate such two-level systems, which are believed to be one of the main
decoherence sources in Josephson devices.

\end{abstract}
\pacs{42.50.Lc, 42.50.Dv, 03.65.Yz} \maketitle

Exploring quantum mechanics in macroscopic artificial quantum
systems consists both theoretical interests and application in
quantum information \cite{Leggett,Makhlin}. Recently, a series of
experiments with devices based on Josephson tunnel junctions (JTJ)
unambiguously demonstrated the
quantum-mechanical behavior of a macroscopic variable \cite%
{Jackel,Nakamura,Yu,Steffen}. The ground state and first excited state, $%
|0\rangle $ and $|1\rangle $ in Fig. 1(a), of the cubic potential well of a
current biased JTJ can be isolated to form an effective two-level
superconducting phase qubit whose state can be manipulated by application of
microwave pulses \cite{Simmonds, Yu}. The superconducting approach is
considered promising for quantum information processing because it is
readily to scale up. However, the realization of superconducting quantum
computer is still hampered by decoherence. It is imperative to investigate
the decoherence mechanism. Recent experiments have shown that JTJs are
ubiquitously coupled to two-level systems (TLS), which are presumably
structural defects within the tunnel junction \cite{Simmonds}. Such TLS has
been identified as one of the dominant sources of decoherence of Josephson
phase qubits. Characterizing the TLS, in particular the effect of an
individual TLS on a Josephson junction, is thus critical to the quest to
improve the performance of superconducting qubits. The first challenge is to
single out a single TLS so that its property can be quantified. The second
more fundamental challenge is that in order to study temporal behavior of an
individual TLS one need to readout its quantum state. In this Letter, we
demonstrate that (1) it is possible to isolate a single TLS from an ensemble
and its coupling to Josephson junction can be controlled via frequency
selection and (2) the phenomenon known as "quantum jump" in such a coupled
TLS-Josephson junction system can be used to readout the quantum state of
the TLS which provides a new approach to calibrate individual TLS and their
interaction with the junction.

The concept of quantum jump came from Bohr who suggested that, the
interaction of light and matter occurs in such a way that an atom undergoes
instantaneous transitions of its internal state upon the emission or
absorption of a light quantum. These sudden transitions have become known as
`quantum jumps' \cite{QJ}. Quantum mechanics is originally considered as a
statistical theory that makes probabilistic predictions of the behavior of
ensembles. Whether it can also be used to describe the dynamics of a single
quantum system has been a subject of debated between some founders of
quantum mechanics \cite{Schrodinger}. Observation of quantum jumps in
isolated atoms and ions unambiguously settled such dispute as quantum jump
is a unique phenomenon displayed only by individual quantum systems that
would be totally masked by averaging over an ensemble \cite{Sauter}. Quantum
jumps in atomic systems also found important applications in quantum
measurement, time and frequency standard, and precision spectroscopy \cite%
{Cook,Blatt}. In contrast, quantum jumps of a macroscopic system is still an
important goal in the field of quantum measurement and its potential
application needs to be explored \cite{Thompson}.

\begin{figure}[tbhp]
\includegraphics[width=7cm,height=7cm]{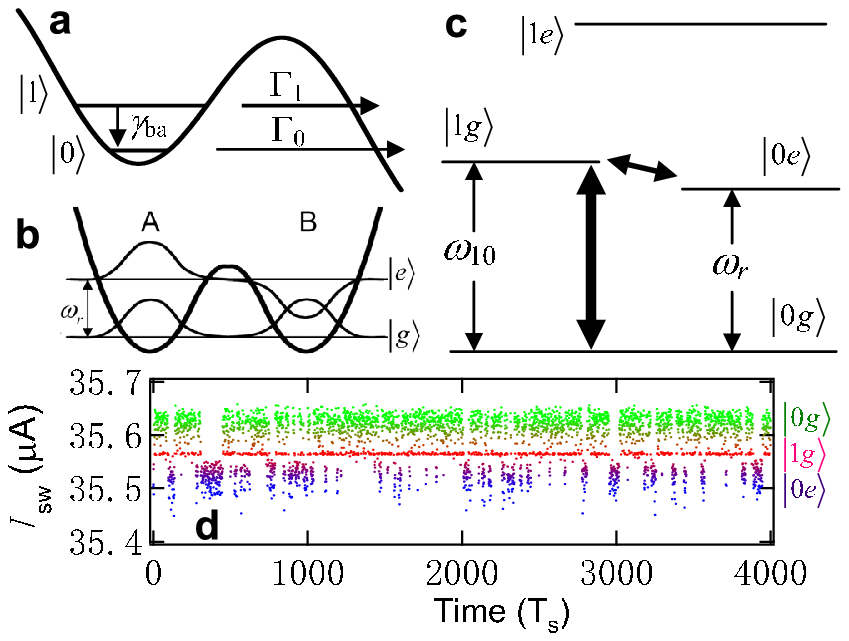}
\caption{ (Color online) (a) Washboard potential of a current
biased JTJ. The relaxation and tunnelling rates are $\gamma _{ba}$, $\Gamma _{0}$, and $%
\Gamma _{1}$. (b) Double well potential for a TLS. $|g\rangle $ and $%
|e\rangle $ represent the ground state and the excited state with
level spacing $\omega _{r}$. (c) Energy level diagram for a
junction coupled to a TLS. The energy levels $|0g\rangle $,
$|1g\rangle $, and $|0e\rangle $ form
the three-level structure used to demonstrate quantum jumps. (d) Simulated $%
I_{sw}$ trajectory with Hamiltonian (1). The time is in the unit
of T$_{s}$, the waveform period of $I_{b}$. Each point represents
$I_{sw}$ obtained in one period of $I_{b}$ cycle.}
\end{figure}

In what follows, we present experimental evidence for quantum jumps of a
single macroscopic quantum system consisting of a current biased JTJ (a
phase qubit) coupled to a microscopic TLS, thus demonstrating a fundamental
quantum effect in a macroscopic device. Since each TLS has a unique
characteristic frequency and the transition frequency $\omega _{10}$ between
the qubit states $|0\rangle $ and $|1\rangle $ is a smooth function of the
bias current $I_{b}$, a qubit can be made to couple to a single TLS to form
a single composite quantum system by adjusting $I_{b}$ of the qubit, so that
its $\omega _{10}$ approaches the level spacing of the target TLS. This
system can be tuned to have an effective three-level structure with a strong
and a weak transition, as depicted in Fig. 1(c), which is similar to the
energy level scheme utilized to observe quantum jumps in atomic systems \cite%
{Blatt}. Because occurrence of the weak transition will cause the strong
transition to turn on and off abruptly we denote the strong transition as
the `on' and weak transition as the `off' state, in analogy to the
terminology used for atomic systems \cite{Blatt}. A direct consequence of
quantum jumps is that the state-dependent switching current $I_{sw}$, at
which the junction switches from zero voltage to finite voltage,\ appears as
a random telegraph signal (RTS) \cite{Cook,Sauter,Blatt}, which is one of
the unambiguous signatures of quantum jumps. As for our experiment, we study
the trajectory (i.e., time record) of $I_{sw}$ of the phase qubit which
showed the predicted RTS when the qubit is irradiated by microwaves with
frequency nearly resonating with the TLS. All key features of quantum jumps
are confirmed in the experiment. We show that coupling between the qubit and
the TLS can be experimentally controlled by adjusting $I_{b}$ and that the
RTS can be used to characterize the property of the TLS. Therefore, the
observation of quantum jumps does not only extend the realm of this quantum
mechanical effects to macroscopic quantum system but also offers an
invaluable tool for the study of decoherence in superconducting qubits.

To describe coupling between the TLS and the phase qubit we adopted a model
proposed by Simmonds \textit{et al.} \cite{Simmonds}. When microwave
radiation with frequency $\omega $ close to the energy level spacing of the
TLS is applied an effective three-level system, as depicted in Fig. 1(c), is
realized. The two states of the TLS correspond to two different
configurations $A$ and $B$ which result in critical currents $I_{c}^{A}$ and
$I_{c}^{B},$ shown in Fig. 1(b). The interaction Hamiltonian between the TLS
and the junction is given by $H_{int}=-\frac{I_{c}^{A}\phi _{0}}{2\pi }\cos
\delta \otimes |\Psi _{A}\rangle \langle \Psi _{A}|-\frac{I_{c}^{B}\phi _{0}%
}{2\pi }\cos \delta \otimes |\Psi _{B}\rangle \langle \Psi _{B}|$, where $%
|\Psi _{A,B}\rangle $ represent the two relavant wave functions of the TLS, $%
\phi _{0}$ is the flux quantum and $\delta $ represents the gauge invarent
phase difference across the junction. If we assume a symmetric potential
with energy eigenstates separated by $\hbar \omega _{r}$, the ground and
excited states are given by $|g\rangle \simeq (|\Psi _{A}\rangle +|\Psi
_{B}\rangle )/\sqrt{2}$ and $|e\rangle \simeq (|\Psi _{A}\rangle -|\Psi
_{B}\rangle )/\sqrt{2}$. Using matrix representation of the phase qubit \cite%
{Martinis}, the Hamiltonian in the basis $\{|a\rangle \equiv |0g\rangle
,|b\rangle \equiv |1g\rangle ,|c\rangle \equiv |0e\rangle \}$ reads

\begin{equation}
H=\hbar \left(
\begin{array}{lll}
-i\Gamma _{a} & \Omega _{c}-\Omega _{m}\cos (\omega t) & 0 \\
\Omega _{c}-\Omega _{m}\cos (\omega t) & \omega _{10}-i\Gamma _{b}-i\gamma
_{ba} & -\Omega _{c} \\
0 & -\Omega _{c} & \omega _{r}-i\Gamma _{c}%
\end{array}%
\right) ,  \label{H_m}
\end{equation}%
where $\Omega _{c}=\frac{(I_{c}^{A}-I_{c}^{B})\phi _{0}\delta _{10}}{4\pi }$%
, $\Omega _{m}=\frac{\phi _{0}\delta _{10}}{2\pi }I_{m}$, $I_{m}$ is the
amplitude of the microwave, and $\delta _{01}=\delta _{10}=\frac{2\pi }{\phi
_{0}}\sqrt{\frac{\hbar }{2\omega _{10}C}}$ is the coupling matrix element, $%
C $ is the capacitance of the junction. Here $\Gamma _{i}$ $(i=a,b,c)$ is
the escape rate from the level $|i\rangle $ out of the potential and $\gamma
_{ba}$ is the rate of energy relaxation from $|b\rangle $ to $|a\rangle $. $%
\Omega _{c}$ is the energy splitting caused by the TLS which can be
determined from the spectroscopy \cite{Simmonds}. Hamiltonian (\ref{H_m})
determines the evolution of the joint state of the phase qubit and the TLS
which can be solved without further approximation through numerical
simulation. However, before presenting the results of simulation, we briefly
discuss the underlying physics that governs the dynamics of the system. In
the interaction picture and choosing a rotating frame of the frequency $%
\omega $, Hamiltonian (\ref{H_m}) can be simplified to

\begin{equation}
H^{\prime }=\hbar \left(
\begin{array}{lll}
-i\Gamma _{a} & -\Omega _{m}/2 & 0 \\
-\Omega _{m}/2 & \Delta -i\Gamma _{b}-i\gamma _{ba} & -\Omega
_{c}e^{-i\Delta t} \\
0 & -\Omega _{c}e^{i\Delta t} & \Delta _{r}-i\Gamma _{c}%
\end{array}%
\right) ,  \label{H_r}
\end{equation}%
where $\Delta \equiv \omega _{10}-\omega $ and $\Delta _{r}\equiv \omega
_{r}-\omega _{10}$ are the detunnings. Disregarding all decay terms the
coupling between the states $|1g\rangle $ and $|0e\rangle $ is the result of
a competition between the diagonal terms $\{\Delta ,\Delta _{r}\}$ and the
off-diagonal term $\Omega _{c}\exp (\pm i\Delta t)$. When the detunnings are
relatively large, this three-level system has a weak transition between the
states $|0e\rangle $ and $|1g\rangle $ and a strong transition between $%
|0g\rangle $ and $|1g\rangle $. This configuration is similar to that used
to observe quantum jumps in atomic systems \cite{Cook,Sauter,Blatt}.

The evolution of the system can be modelled using a standard Monte Carlo
simulation developed specifically for quantum jumps \cite{Blatt}. In the
simulation, in accordance with the experimental procedure\ we sweep the bias
current $I_{b}$\ linearly\ from a initial value $<<I_{c}$ to a value
slightly larger than $I_{c}$\ within time $T_{0}$. We divide $T_{0}$\ into N
intervals for sufficiently\ large $N$\ and calculate the evolution
trajectory of the coupled system described by Eq.(1). In each interval, we
compare the probability $p$\ of the qubit switching to the finite voltage
state\ within the\ interval with a random number $r$\ uniformly distributed\
between zero and one. A switching\ event is assumed to occur if $p>r$\ and
the bias current at this interval is then the switching\ current $I_{sw}$.
Otherwise the process is repeated\ in the next interval until a switching\
event finally occurs.  Figure 1(d) shows a simulated $I_{sw}$ trajectory
using parameters appropriate for our qubit. $I_{sw}$ is clearly clustered
around three values. The upper, middle, and lower values correspond to the
junction escaped to the finite voltage state from the state $|0g\rangle $, $%
|1g\rangle $ and $|0e\rangle $, respectively. Since the coupling between $%
|0g\rangle $\ and $|1g\rangle $\ is much stronger than that between $%
|1g\rangle $\ and $|0e\rangle $\ the system undergoes frequent transitions
between $|0g\rangle $\ and $|1g\rangle $\ and infrequent jumps between $%
|1g\rangle $\ and $|0e\rangle $. The abrupt changes in $I_{sw}$ of the qubit
mark the occurrence of quantum jumps between the three levels of the system.
The parameters used in the simulation are: $\omega /2\pi =\omega _{10}/2\pi
=9.02$ GHz, $\omega _{r}/2\pi =8.7$ GHz, $\Omega _{c}/2\pi =200$ MHz (Fig.
2(c)), $\Omega _{m}/2\pi \sim 2$ MHz, and $\gamma _{ba}=0.6\mu s^{-1}$ \cite%
{sun07}. All parameters, except coupling strength between the junction and
the microwave field $\Omega _{m}$, were determined from independent
measurements and also consistent with the previous reports \cite{Yu}. From
the simulation we found that $\Omega _{m}/2\pi \sim 2$ MHz, which is
consistent with our experimental setup.

\begin{figure}[tbhp]
\includegraphics[width=7cm,height=7cm]{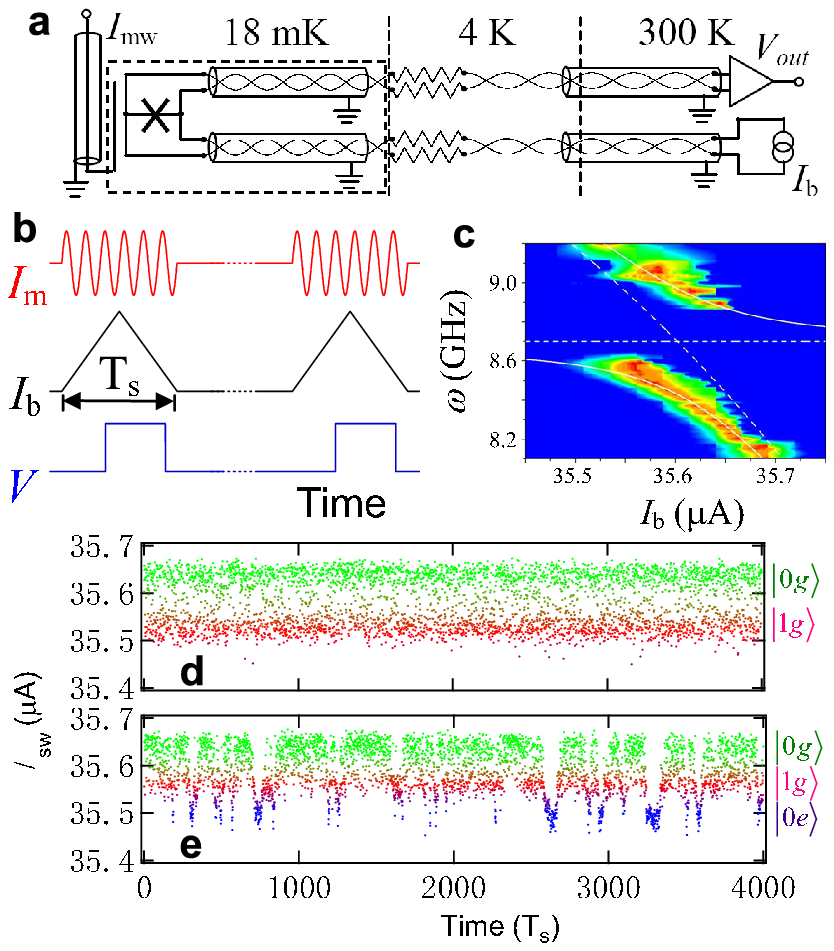}
\caption{(Color online) (a) Schematic drawing of the measurement
circuit. X represent a junction. (b) The time profile of microwave
signal, bias current, and voltage signals. (c) Avoided crossing on
the qubit spectroscopy caused by the coupling between the qubit
and TLS. (d) $I_{sw}$ trajectory with off resonant, $\omega /2\pi
=9.2\ \text{GHz}>\omega _{r}/2\pi $, and (e) on resonant, $\omega
/2\pi =9.02$ GHz, microwave irradiation at 18 mK. }
\end{figure}

Experimentally, we implemented the three-level system using a $10$ $\mu $m$%
\times 10$ $\mu $m Nb/AlO$_{x}$/Nb junction. The parameters of the junction
are $I_{c}$ $\approx 36\ \mu \text{A}$ and $C\approx 4$ pF, respectively.
The device was thermally anchored to the mixing chamber of a dilution
refrigerator with a base temperature of about $18$ mK. \textbf{\ }As
schematically shown in Fig. 2(a), the switching current of the junction was
measured by the four-probe method. The measurement lines were composed of
twisted wires, RC filters, and copper powder filters \cite{Yu07}. The center
conductor of an open-ended coaxial cable was placed above the junction for
application of microwave. This arrangement resulted in $>$110 dB attenuation
between the end of the coaxial cable and the junction. A saw-tooth bias
current was applied with a reputation rate of 100 Hz (Fig.2 (b)). The
junction voltage was amplified by a differential amplifier and the switching
current was recorded when a voltage greater than the threshold was first
detected during every ramp. In the experiment, we first spectroscopically
examined the phase qubit by measuring $\omega _{10}$ as a function of $I_{b}$
(Fig. 2(c)) in the region slightly below $I_{c}$. The measured $\omega
_{10}(I_{b})$ agreed well with the theoretical prediction of the $|0\rangle $
to $|1\rangle $ transition \cite{Yu,Simmonds}. An energy-level repulsion
caused by coupling to a single TLS in the junction was observed in the
spectrum when the frequency of the microwave $\omega /2\pi $ was about $8.7$
GHz. In the vicinity of this anticrossing the TLS and qubit formed an
effective three-level system which was utilized to observe the
aforementioned quantum jumps. The quantum states of the system were
determined from the values of $I_{sw}$.

Figures 2(d) and 2(e) show the measured $I_{sw}$ trajectory as a function of
time. When the microwave source is not resonant with the level spacing of
the TLS (Fig. 2(d)) the junction and TLS are effectively decoupled. However,
the interaction between the junction and the microwave field could still
generate transitions between the qubit states $|0\rangle $ and $|1\rangle $
. In this case the junction could tunnel out of the potential well either
from $|0\rangle $ or $|1\rangle $. Note that $I_{sw}$ of $|0\rangle $ state
is slightly higher than that of $|1\rangle $ state. In Fig. 2(d) the
trajectories are clustered around $35.63\ \mu \text{A}$ and $35.55\ \mu
\text{A}$, corresponding to $I_{sw}$ for qubit states $|0\rangle $ and $%
|1\rangle $ respectively. Here, only the strong transition between $%
|0g\rangle $ and $|1g\rangle $ was observed because only these two levels
were involved in the dynamics of the system. When a microwave field with $%
\omega /2\pi =9.02$ GHz was applied (Fig. 2(e)) coupling between the qubit
and TLS was turned on. In this case $I_{sw}$ trajectory jumped randomly
between three distinctive bands: the upper $(35.63\ \mu \text{A})$, middle $%
(35.55\ \mu \text{A})$, and lower $(35.50\ \mu \text{A})$ bands,
corresponding to the coupled qubit-TLS system being in the state $|0g\rangle
$, $|1g\rangle $, and $|0e\rangle $, respectively \cite{Endnote}. Such RTS
is the fingerprint of quantum jumps.

One of the prominent signatures of quantum jumps is the exponential decay of
the dwell time between the jumps \cite{Blatt}. To analyze our data we borrow
from the general methods used to study quantum jumps in atomic systems. We
hereafter group the two states $\{|0g\rangle ,|1g\rangle \}$ coupled by the
strong transition as one state (denoted as `on') and $|0e\rangle $ as the
other state (denoted as `off'). Then the system can be fully characterized
by the dwell times in the `on' and `off' states and the upward (downward)
transition rate $R_{on}$ ($R_{off}$). We extracted distributions of the `on'
and `off' time intervals from the experimental data. In each case the
results were binned and fit to exponential decays (Fig. 3(a), 3(b)). The
transition rates for the `on' and `off' states obtained from the best fits
are $R_{on}=0.236\pm 0.036\ s^{-1}$ and $R_{off}=2.38\pm 0.45\ s^{-1}$,
respectively. The uncertainties of $R$s were determined following the
standard procedure with $90\%$ confidence level. Note that $R_{off}$ is the
relaxation rate of the state $|e\rangle $ of the TLS, an important property
that otherwise is very difficult to measure. Thus the method presented here
enables further investigation of $R_{off}$ as a function of relevant
parameters,\ which is crucial to understand the role of TLS in qubit
decoherence.

\begin{figure}[tbhp]
\includegraphics[width=7cm,height=5cm]{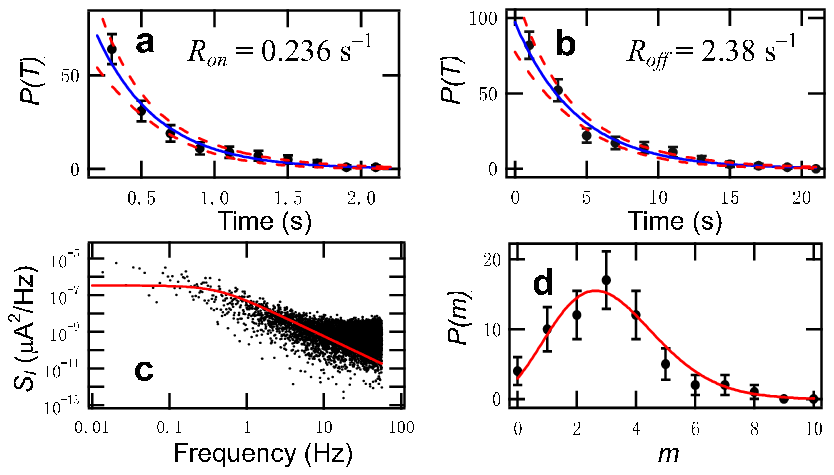}
\caption{(Color online) The distribution of the length of (a) `on'
time intervals and (b) `off' time intervals obtained from
experimental data.\ The solid lines are best fits to exponential
decay and the dashed lines indicate the corresponding 90\%
confidence bands.\ (c) Power spectrum of $I_{sw}$ trajectory at
$\omega /2\pi =9.02$ GHz. Solid line was calculated by
inserting $R_{on}$ ($R_{off})$ obtained from the fits in 3(a) (3(b)) into $%
S_{I}(f)$ equation and scaling by $\kappa ^{2}$. (d) The
distribution of `off' to `on' jumps. $m$ is the events of `off' to
`on' jumps in 6 s. The solid line is a fit to a Poisson
distribution. }
\end{figure}

It is also predicted \cite{Cook} that the power spectrum of $I_{sw}$
trajectory is a Lorentzian with half-width $2\pi \Delta f=R_{on}+R_{off}$, $%
S_{I}(f)=\frac{\kappa ^{2}R_{on}R_{off}}{R_{on}+R_{off}}\frac{1}{(2\pi
f)^{2}+(R_{on}+R_{off})^{2}}$, where $\kappa $ has the dimension of current.
To verify that the observed RTS is indeed due to coupling to a sigle TLS we
digitized $I_{sw}$ in the `on' and `off' states to obtain its power spectrum
using Fourier transformion. As shown in Fig. 3 (c), the shape of the power
spectrum of RTS agrees with the theoretical result well.

Another verifiable feature of quantum jumps is the functional form of jump
history versus time. It is predicted that the distribution of `off' to `on'
jumps follows the Poisson statistics \cite{Blatt}. This Poisson distribution
is confirmed by the data presented in Fig. 3(d). Therefore, both experiments
and simulations showed that $I_{sw}$ trajectory jumps exactly between three
distinctive states. Since these macroscopic quantum states represent the
motion of about $10^{9}$ electrons, this suggests that $10^{9}$ electrons
jump simultaneously to other state in absorption or emission of a single
photon. We emphasize that when temperature was increased to $0.2$ K, which
is above the quantum-to-classical crossover temperature, RTS in $I_{sw}$
disappeared. Hence, our result cannot be explained by a non-quantum
mechanical mechanism.

We thank J. Cao, Y. Wang, W. Xu, L. Kang for technical assistance and R.
Harris, D. Xing for thoughtful comments. This work is partially supported by
NSFC, NBRP of China (2006CB921801 and 2006CB601006). S. H. acknowledges
support by NSF (DMR-0325551).



\begin{thebibliography}{99}



\bibitem{Leggett} A. J. Leggett and A. Garg, Phys. Rev. Lett. \textbf{54},
857 (1985); W. H. Zurek, Phys. Today \textbf{44}, 36 (1991).

\bibitem{Makhlin} Y. Makhlin, G. Sch\"{o}n, and A. Shnirman, Rev. Mod. Phys.
\textbf{73}, 357 (2001).

\bibitem{Jackel} L. D. Jackel \textit{et al.}, Phys. Rev. Lett. \textbf{47},
697 (1981); R. F. Voss and R. A. Webb, \textit{ibid.}, 265 (1981); R. H.
Koch, D. J. Van Harlingen, and J. Clarke, Phys. Rev. Lett. \textbf{47}, 1216
(1981); J. Clarke \textit{et al.}, Science \textbf{239}, 992 (1988); J. M.
Martinis \textit{et al.,} Phys. Rev. B \textbf{35}, 4682 (1987).

\bibitem{Nakamura} Y. Nakamura, Y. A. Pushkin, and J. S. Tsai, Nature
\textbf{398}, 786 (1999); D. Vion \textit{et al.}, Science \textbf{296}, 886
(2002); I. Chiorescu \textit{et al.}, Science \textbf{299}, 1869 (2003).

\bibitem{Yu} Y. Yu \textit{et al.}, Science \textbf{296}, 889 (2002); R.
McDermott \textit{et al.}, Science \textbf{307}, 1299 (2005).

\bibitem{Steffen} M. Steffen \textit{et al.}, Science \textbf{313}, 1423
(2006); A. Wallraff \textit{et al.}, Nature \textbf{431}, 162 (2004); I.
Chiorescu \textit{et al.}, Nature \textbf{431}, 159 (2004); J. Mayer \textit{%
et al.}, Nature \textbf{449}, 443 (2007); M. A. Sillanpaa, J. I. Park, and
R. W. Simmonds, Nature \textbf{449}, 438 (2007).

\bibitem{Simmonds} R. W. Simmonds \textit{et al.}, Phys. Rev. Lett. \textbf{%
93}, 077003 (2004).

\bibitem{QJ} M.O.Scully and M.S.Zubariry, \textit{Quantum optics}
(Cambridge, 1997).

\bibitem{Schrodinger} E. Schr\"{o}dinger, Br. J. Phil. Sci. III \textbf{10},
109 (1952).

\bibitem{Sauter} Th. Sauter \textit{et al.}, Phys. Rev. Lett. \textbf{57},
1696 (1986); J. C. Bergquist \textit{et al.}, Phys. Rev. Lett. \textbf{57},
1699 (1986); T. Erber and S. Putterman, Nature \textbf{318}, 41 (1985).

\bibitem{Cook} H. Dehmelt, Bull. Am. Phys. Soc. \textbf{20}, 60 (1975); R.
J. Cook and H. J. Kimble, Phys. Rev. Lett. \textbf{54}, 1023 (1985).

\bibitem{Blatt} R. Blatt and P. Zoller, Eur. J. Phys. \textbf{9}, 250
(1988); M. B. Plenio and P. L. Knight, Rev. Mod. Phys. \textbf{70}, 101
(1998).

\bibitem{Thompson} J. D. Thompson \textit{et al.}, Nature \textbf{452}, 72
(2008).

\bibitem{Martinis} J. M. Martinis \textit{et al.}, Phys. Rev. B \textbf{67},
094510 (2003).

\bibitem{sun07} G. Sun \textit{et al.}, Supercond. Sci. Tech. \textbf{20},
S437 (2007).

\bibitem{Yu07} Y. Yu \textit{et al.}, Supercond. Sci. Tech. \textbf{20},
S441 (2007).

\bibitem{Endnote} When a single TLS is coupled to a phase qubit the junction
will have two distinctive critical currents $I_{c,g}$ and $I_{c,e}$,
corresponding to the TLS being in the ground and excited states,
respectively. If coupling between the qubit and TLS is sufficiently strong
one can have $I_{sw,0g}-I_{sw,0e}\simeq $ $%
I_{c,g}-I_{c,e}>I_{sw,0g}-I_{sw,1g}$, thus $I_{sw,1g}>$ $I_{sw,0e}$ as the
case in our experiment.
\end{thebibliography}
\end{document}